\documentclass[a4paper,reqno]{amsart}

\usepackage[all]{xy}           %For commutative diagrams
\usepackage{amssymb}           %For double arrows
\usepackage{hyperref}
\usepackage{eucal}
\numberwithin{equation}{section}

\newtheorem{definition}{Definition}[section]

\newtheorem{theorem}[definition]{Theorem}

\newtheorem{corollary}[definition]{Corollary}
\newtheorem{remarkth}[definition]{Remark}

\newenvironment{remark}{\begin{remarkth}\upshape}{\hfill$\diamond$\end{remarkth}}

\renewcommand{\emph}[1]{{\bfseries\itshape{#1}}}

\newcommand{\R}{\mathbb{R}}      %Numeros reales
\newcommand{\ds}{\displaystyle}

\makeatletter
\newcommand\prol{\@ifstar{\@proldf}{\@prolpf}}  %% if * dual else primal
\def\@prolpf{\@ifnextchar[{\@prolpf@wrt}{\@prolpf@}}
\def\@prolpf@wrt[#1]#2{\@ifnextchar[{\@prolpf@wrt@at{#1}{#2}}{\@prolpf@wrt@{#1}{#2}}}
\def\@prolpf@wrt@at#1#2[#3]{\prolsymbol^{#1}_{#3}#2}
\def\@prolpf@wrt@#1#2{\prolsymbol^{#1}#2}
\def\@prolpf@#1{\@ifnextchar[{\@prolpf@at{#1}}{\@prolpf@@{#1}}}
\def\@prolpf@at#1[#2]{\prolsymbol_{#2}#1}
\def\@prolpf@@#1{\prolsymbol#1}
\def\@proldf{\@ifnextchar[{\@proldf@wrt}{\@proldf@}}
\def\@proldf@wrt[#1]#2{\@ifnextchar[{\@proldf@wrt@at{#1}{#2}}{\@proldf@wrt@{#1}{#2}}}
\def\@proldf@wrt@at#1#2[#3]{\prolsymbol^{*#1}_{#3}#2}
\def\@proldf@wrt@#1#2{\prolsymbol^{*#1}#2}
\def\@proldf@#1{\@ifnextchar[{\@proldf@at{#1}}{\@proldf@@{#1}}}
\def\@proldf@at#1[#2]{\prolsymbol^*_{#2}#1}
\def\@proldf@@#1{\prolsymbol^*#1}
\def\prolsymbol{\mathcal{T}}
\makeatother

\begin{document}

\title[Hamilton-Jacobi theory in $k$-symplectic field theories]{Hamilton-Jacobi theory in $k$-symplectic field theories}

\author[M. de Le\'on]{M. de Le\'on}
\address{M. de Le\'on:
Instituto de Ciencias Matem\'aticas (CSIC-UAM-UC3M-UCM),
Consejo Superior de Investigaciones Cient\'{\i}ficas, Serrano 123, 28006
Madrid, Spain} \email{mdeleon@icmat.es}

\author[D.\ Mart\'{\i}n de Diego]{D. Mart\'{\i}n de Diego}
\address{D.\ Mart\'{\i}n de Diego:
Instituto de Ciencias Matem\'aticas (CSIC-UAM-UCM-UC3M), Consejo
Superior de Investigaciones Cient\'{\i}ficas\\ C/ Serrano 123, 28006
Madrid, Spain} \email{david.martin@icmat.es}

\author[J.C. Marrero]{J.C. Marrero}
\address{J.C. Marrero: Departamento de Matem\'{a}tica Fundamental, Facultad de Matem\'{a}ticas, Universidad de la Laguna, Spain}
\email{jcmarrer@ull.es}

\author[M. Salgado]{M. Salgado}
\address{M. Salgado:
Departamento de Xeometr\'{\i}a e Topolox\'{\i}a, Facultade de Matem\'{a}ticas,
    Universidade de Santiago de Compostela,
    15782-Santiago de Compostela, Spain}
\email{modesto.salgado@usc.es}

\author[S. Vilari\~no]{S. Vilari\~no}
\address{S. Vilari\~no:
Departamento de Matem\'{a}ticas, Facultad de Ciencias,
    Universidad de A Coru\~{n}a,
    Campus de A Zapateira,
    15008-A Coru\~{n}a, Spain}
    \email{silvia.vilarino@udc.es}

\today{ }

\keywords{Hamilton-Jacobi theory, $k$-symplectic field theories.}

 \subjclass[2000]{}

\begin{abstract}
 In this paper we extend the geometric formalism of Hamilton-Jacobi
 theory for Mechanics to the case of classical field theories in the $k$-symplectic framework.

\end{abstract}

%\thanks{This work has been partially supported by MICYT (Spain)
%Grant MTM 2006-7832 and...}

 \maketitle

\tableofcontents

\section{Introduction}

The usefulness of Hamilton-Jacobi theory in Classical Mechanics
is well-known, giving an alternative  procedure to study and, in some  cases, to solve the evolution equations \cite{am}.
The use of symplectic geometry in the study of Classical Mechanics
has permitted to connect the Hamilton-Jacobi theory with the theory
of lagrangian submanifolds and generating functions.

At the beginning of the 1900s an analog of Hamilton-Jacobi equation for field theory has been developed
\cite{rund}, but it has not been proved to be as powerful as the theory which is available for mechanics \cite{bertin,bruno,pau1,pau2,rosen,vita}.

Our goal in this paper is to describe this equation
in a geometrical setting.

Let us recall that there are two different ways to describe a field theory, say
multisymplectic  and $k$-symplectic geometry.
A multisymplectic structure abstracts the canonical geometry of bundles of exterior forms,
in the same way that symplectic geometry captures the essential facts on cotangent bundles \cite{CIL1}.
On the contrary, a $k$-symplectic structure is locally equivalent to the Whitney sum $(T^1_k)^*Q = T^*Q \oplus \stackrel{k}{\cdots} \oplus T^*Q$
of $k$-copies of the cotangent bundle $T^*Q$. In any case, given a Hamiltonian function, both geometric structures
produce the field equations.

The aim of this paper is to extend the Hamilton-Jacobi theory to field theories just in
the context of $k$-symplectic manifolds (we remit to \cite{lmm2}
for a description in the multisymplectic setting). The dynamics for a given hamiltonian function $H$ is interpreted
as a family of vector fields (a $k$-vector field) on the phase space $(T^1_k)^*Q$.
The Hamilton-Jacobi equation is of the form
$$
d(H \circ \gamma) = 0,
$$
where $\gamma = (\gamma_1, \dots, \gamma_k)$ is a family of closed
$1$-forms on $Q$. Therefore, we recover the classical form
$$
H(q^i,\ds\frac{\partial W^1}{\partial q^i}, \ldots
,\ds\frac{\partial W^k}{\partial q^i}) = constant\;.
$$
where $\gamma_i = dW_i$.
It should be noticed that our method is inspired in a  recent result by Cari\~{n}ena {\it et al} \cite{CGMMR} (this method
has also used to develop a Hamilton-Jacobi theory for nonholonomic mechanical systems \cite{lmm1}; see also \cite{lmm3,pepin2}).

The paper is structured as follows. In Section 2, we recall the notion of $k$-vector field and their
integral sections. In Section 3 we discuss $k$-symplectic Hamiltonian field theory and the Hamilton-Jacobi
equation in that context. The corresponding result in the lagrangian description of the field theory
is obtained in Section 4.
Finally, an example is discussed in Section 5, with the aim to show how the method
works.

\section{Geometric preliminaires}

In this section we briefly recall some well-known facts about
tangent bundles of $k^1$-velocities (we refer the reader to
\cite{mt1,mt2,arxive,mor,MRS-2004} for more details).

Let $\tau_M : TM \longrightarrow M$ be the tangent bundle of $M$.
Let us denote by $T^1_kM$ the Whitney sum $TM \oplus
\stackrel{k}{\dots} \oplus TM$ of $k$ copies of $TM$, with
projection $\tau: T^1_kM \longrightarrow M$, $\tau
({v_1}_{x},\dots , {v_k}_{x})=x$, where ${v_A}_{x}\in T_{x}M$,
$1\leq A \leq k$. $T^1_kM$ can be identified with the manifold
$J^1_0(\R^k,M)$ of the $k^1$-velocities of $M$, that is, $1$-jets of
maps $\eta : \R^k \longrightarrow M$ with source at $0 \in \R^k$,
say
\[
\begin{array}{ccc}
J^1_0(\R^k,M) & \equiv & TM \oplus \stackrel{k}{\dots} \oplus TM \\
j^1_{0, x} \eta & \equiv & ({v_1}_x, \dots , {v_k}_x)
\end{array}
\]
where $x=\eta (0)$, and ${v_A}_x = T\eta
(0)(\ds\frac{\partial}{\partial t^A} \Big\vert_{0})$. Here
$(t^1,\ldots, t^k)$ denote the standard coordinates on $\R^k$.
$T^1_kM$ is called the {\it tangent bundle of $k^1$-velocities} of
$M$ or simply $k$-tangent bundle for short, see \cite{mor}.

Denote by $(x^i , v^i)$ the fibred coordinates in $TM$ from local
coordinates $(x^i)$ on $M$. Then we have fibred coordinates $(x^i ,
v_A^i)$, $1\leq i \leq m,\, 1 \leq A \leq k$, on $T^1_kM$, where
$m=\dim M$.

\begin{definition}\label{kvector}
A section ${\bf X} : M \longrightarrow T^1_kM$ of the projection
$\tau$ will be called a {\rm $k$-vector field on $M$}.
\end{definition}

Since $T^{1}_{k}M$ is  the Whitney sum $TM\oplus \stackrel{k}{\dots}
\oplus TM$ of $k$ copies of $TM$, we deduce that to give a
$k$-vector field ${\bf X}$ is equivalent to give a family of $k$
vector fields $X_{1}, \dots, X_{k}$ on $M$ by projecting ${\bf X}$
onto each factor. For this reason we will denote a $k$-vector field
by $(X_1, \dots, X_k)$.

\begin{definition}
\label{integsect} {\rm An integral section} of the $k$-vector field ${\bf
X}=(X_{1}, \dots,X_{k})$, passing through a point $x \in M$, is a
map $\psi \colon U_0 \subset \R^k \longrightarrow M$, defined on
some neighborhood $U_0$ of $0 \in \R^k$, such that
$$
\psi(0) = x, \, \, T\psi\left(\ds\frac{\partial}{\partial
t^A}\Big\vert_t\right)=X_{A}(\psi (t))
 \quad , \quad \mbox{\rm for every $t \in U_0$, $1\leq A \leq k$}
$$
or, what is equivalent,  $\psi$ satisfies that
$X\circ\psi=\psi^{(1)}$, being  $\psi^{(1)}$ is the first
prolongation of $\psi$  to $T^1_kM$ defined by
$$
\begin{array}{rccl}\label{1prolong}
\psi^{(1)} : & U_0\subset \R^k & \longrightarrow & T^1_kM
\\\noalign{\medskip}
 & t & \longrightarrow & \psi^{(1)} (t)= j^1_0 \psi_t \;,
 \end{array}
$$
where $\psi_t(s) = \psi(t+s)$.

A $k$-vector field ${\bf X}=(X_1,\ldots , X_k)$ on $M$ is said to be
integrable if there is an integral section passing through every
point of $M$.
\end{definition}

In local coordinates, we have
\begin{equation} \label{localfi11}
\psi^{(1)}(t^1, \dots, t^k)=\left( \psi^i (t^1, \dots, t^k),
\ds\frac{\partial\psi^i}{\partial t^A} (t^1, \dots, t^k)\right), \,
1\leq A\leq k,\,  1\leq i\leq m \, ,
\end{equation}
and then $\psi$ is an integral section of $(X_1, \dots, X_k)$ if and
only if the following equations holds:
$$
\ds\frac{\partial \psi^i}{\partial t^A}= X_A^i \circ \psi\, \quad
1\leq A \leq k,\; 1\leq i\leq m\;,
$$
being $X_A=X_A^i\ds\frac{\partial}{\partial q^i}$.

Notice that, in case $k=1$, Definition \ref{1prolong} coincides with
the definition of integral curve of a vector field.

\section{$k$-symplectic Hamiltonian field theory and the
Hamilton-Jacobi equation}

In this section, we shall recall the $k$-symplectic Hamiltonian
formulation for classical field theories, (see \cite{gun,MRS-2004}
for more details). Later,  we shall describe the Hamilton-Jacobi
problem in this setting.

\subsection{$k$-symplectic Hamiltonian
field theory}

Let $Q$ be a configuration manifold with local coordinates $(q^i)$, $1\leq i\leq n$ and $T^*Q$ its cotangent bundle with fibered coordinates $(q^i, p_i)$.
Denote by $\pi_Q: T^*Q\to Q$ the canonical projection.
Define the Liouville 1-form or canonical 1-form $\theta_Q$ by
\[
(\theta_Q)_{\alpha}(Y)=\alpha(T\pi_Q(Y)), \hbox{  where  } Y\in T_{\alpha}(T^*Q)
\]
The canonical 2-form $\omega_Q$ on $T^*Q$ is the symplectic form $\omega_Q=-d\theta_Q$.
Therefore, we have
\begin{equation}\label{symp}
\theta_Q=p_i\, dq^i\; ,\qquad \omega_Q = dq^i \wedge dp_i\;.
\end{equation}

Denote
$$
(T^1_k)^*Q = T^*Q \oplus \stackrel{k}{\cdots} \oplus T^*Q
$$
the Whitney sum of $T^*Q$ with itself $k$ times. We introduce
coordinates $(q^i, p^1_i, \dots, \linebreak p^k_i)$ and the
canonical projections
$$
\Pi_Q : (T^1_k)^*Q \longrightarrow Q \; , \; \Pi_A : (T^1_k)^*Q
\longrightarrow T^*Q
$$
where $A$ indicates the summand $A$-th in the Whitney sum.

We can endow $(T^1_k)^*Q$ with a $k$-symplectic structure given by
the family of $k$ canonical presymplectic forms $(\omega^1, \dots,
\omega^k)$, where
$$
\omega^A = \Pi_A^* (\omega_Q)\;.
$$
Therefore, we have
$$
\omega^A = dq^i \wedge dp^A_i\;.
$$
Denote also by $\theta^A = \Pi_A^* (\theta_Q)$.

\begin{remark}
Let us recall that a $k$-symplectic structures is given by a family of $k$ two-forms
satisfying some compatibility conditions (see \cite{aw1,aw3,merino1,merino2,merino3,arxive}).
\end{remark}

Consider a Hamiltonian $H : (T^1_k)^*Q \longrightarrow \R$.
The field equations are then obtained as follows.

Consider the mapping
\begin{eqnarray*}
\flat &:& T_k^1((T^1_k)^*Q) \longrightarrow T^*((T^1_k)^*Q)\\
&& z = (z_1, \dots, z_k)  \mapsto \flat(z) = \hbox{trace} \left(
i_{z_A} \, \omega^B\right)=\sum_{A=1}^ki_{z_A}\omega^A
\end{eqnarray*}

Then, we look for the solutions of the equation
\begin{equation}\label{hameq}
\flat(Z) = dH\;.
\end{equation}

Notice that $Z=(Z_1,\ldots, Z_k)$ is a $k$-vector field on $(T^1_k)^*Q$, that is, each $Z_A$ is a vector field on $(T^1_k)^*Q$.

Using a local coordinates system $(q^i,p^A_i)$ on $(T^1_k)^*Q$, each $Z_A$ is locally given by $$
Z_A = Z_A^i \ds\frac{\partial}{\partial q^i} + (Z_A)^B_i
\ds\frac{\partial}{\partial p^B_i }\; .
$$

Therefore, we obtain that the equation (\ref{hameq}) is locally expressed as follows:
\begin{equation}\label{localhameq}
Z_A^i = \ds\frac{\partial H}{\partial p^A_i} \quad , \quad
\ds\sum_{A=1}^k(Z_A)^A_i = - \ds\frac{\partial H}{\partial q^i}\;.
\end{equation}

Now, if $Z$ is integrable, an integral section of $Z$
$$
\sigma(t^1, \dots, t^k) = (\sigma^i(t^1, \dots, t^k),
\sigma^A_i(t^1, \dots, t^k))
$$
satisfies the Hamilton equations
\begin{equation}\label{localhameq2}
\ds\frac{\partial \sigma^i}{\partial t^A}  = \ds\frac{\partial H}{\partial
p^A_i}\circ\sigma \quad  , \quad   \ds\sum_{A=1}^k\ds\frac{\partial \sigma^A_i}{\partial t^A} = -
\ds\frac{\partial H}{\partial q^i}\circ\sigma\;.
\end{equation}

 Let us observe that  if $Z = (Z_1, \dots, Z_k)\in \ker \flat$ then
\begin{equation}\label{kerb}
Z_B^i=0 \quad, \quad \ds\sum_{A=1}^k (Z_A)^A_i=0 \,  .
\end{equation}

\subsection{The Hamilton-Jacobi equation}\label{HJproblem}

The standard formulation of the Hamilton-Jacobi problem for Hamiltonian Mechanics consist of finding a function $S(t,q^i)$ (called the {\it principal function}) such that
\begin{equation}\label{H-JeqHmech}
\ds\frac{\partial S}{\partial t} + H(q^i,\ds\frac{\partial
S}{\partial q^j})=0\,.
\end{equation}

If we put $S(t,q^i)= W(q^i)-t \cdot  constant$, then $W\colon Q\to \R$ (called the {\it characteristic function}) satisfies
\begin{equation}\label{H-JeqHmech2}
  H(q^i,\ds\frac{\partial W}{\partial q^j})=constant\,.
\end{equation}

Equations (\ref{H-JeqHmech}) and (\ref{H-JeqHmech2}) are indistinctly referred as the {\it Hamilton-Jacobi equation} in Hamiltonian Mechanics.

In the framework of the $k$-symplectic formalism, a Hamiltonian is a
function $H\in\mathcal{C}^\infty((T^1_k)^*Q)$. In this context, the
Hamilton-Jacobi problem consists of finding $k$ functions
$W^1,\ldots, W^k\colon Q\to \R$ such that
\begin{equation}\label{HJ-ksym}
H(q^i,\ds\frac{\partial W^1}{\partial q^i}, \ldots
,\ds\frac{\partial W^k}{\partial q^i}) = constant\;.
\end{equation}

In this section we give a geometric version of the Hamilton-Jacobi equation (\ref{HJ-ksym}).

Let $\gamma : Q \longrightarrow (T^1_k)^*Q$ be a closed section of $\Pi_Q :
(T^1_k)^*Q \longrightarrow Q$. Therefore, $\gamma = (\gamma^1,
\dots, \gamma^k)$ where each $\gamma^A$ is an ordinary closed 1-form on
$Q$. Thus we have that
every point has an open neighborhood $U\subset Q$ where there exists
$k$ functions $W^A\in\mathcal{C}^\infty(U)$ such that $\gamma^A=dW^A$.

Now, let $Z$ be a $k$-vector field on $(T^1_k)^*Q$. Using $\gamma$ we
can construct a $k$-vector field $Z^\gamma$ on $Q$ such that the following
diagram is commutative

\[
 \xymatrix{ (T^1_k)^*Q
\ar[dd]^{\Pi_Q} \ar[rrr]^{Z}&   & &T_k^1((T^1_k)^*Q)\ar[dd]^{T^1_k\Pi_Q}\\
  &  & &\\
 Q\ar@/^1pc/[uu]^{\gamma}\ar[rrr]^{Z^{\gamma}}&  & & T_k^1Q }
\]
that is,
$$Z^\gamma:= T^1_k\Pi_Q\circ Z\circ \gamma\;.$$
Let us remember that for an arbitrary differentiable map $f:N\to M$, the induced
map $T^1_kf:T^1_kN\to T^1_kM$ is defined by $T^1_kf({v_1}_x, \dots ,
{v_k}_x)= (T_xf({v_1}_x),\ldots,T_xf({v_k}_x))$.

 Notice that the $k$-vector field $Z$ defines $k$ vector fields on $(T^1_k)^*Q$, say $Z
= (Z_1, \dots, Z_k)$. In the same manner, the $k$-vector field
$Z^\gamma$ determines $k$ vector fields on $Q$, say $Z^\gamma =
(Z^\gamma_1, \dots, Z^\gamma_k)$.

In local coordinates, if each $Z_A$ is locally given by
$$
Z_A = Z^i_A \, \ds\frac{\partial}{\partial q^i} + (Z_A)^B_i\ds\frac{\partial}{\partial p^B_i}\,,
$$ then $Z^\gamma_A$ has the following local expression:
\begin{equation}\label{zgamma}
 Z^\gamma_A = (Z_A^i \circ\gamma) \, \ds\frac{\partial}{\partial q^i}\,.
\end{equation}

\begin{theorem}\label{hjth} (Hamilton-Jacobi Theorem)
Let $Z$ be a solution of the Hamilton equations (\ref{hameq}) and
$\gamma : Q \longrightarrow (T^1_k)^*Q$ be a   closed section  of
$\Pi_Q : (T^1_k)^*Q \longrightarrow Q$, that is, $\gamma =
(\gamma^1, \dots, \gamma^k)$ where   each $\gamma^A$ is an ordinary
closed 1-form on $Q$. If $Z$ is integrable then the following
statements are equivalent:
\begin{enumerate}
\item If $\sigma\colon U\subset \R^k\to Q$ is an integral section of $Z^\gamma$  then $\gamma\circ\sigma$ is a solution of the Hamilton equations;

\item $d(H\circ \gamma)=0$.

\end{enumerate}
\end{theorem}
 \proof   The closeness of the $1$-forms $\gamma^A=\gamma^A_idq^i$ states that
 \begin{equation}\label{closeness}
 \ds\frac{\partial \gamma^B_i}{\partial q^j} = \ds \frac{\partial \gamma^B_j}{\partial q^i}\,.
 \end{equation}

 \noindent $(i)\Rightarrow (ii)$

 Let us suppose that  $\gamma\circ \sigma(t)=(\sigma^i(t),\gamma^A_i(\sigma(t)))$ is a solution of the Hamilton equations for $H$, then
 \begin{equation}\label{hegamma}
 \ds\frac{\partial \sigma^i}{\partial t^A}\Big\vert_{t}=\ds\frac{\partial H}{\partial p^A_i}\Big\vert_{\gamma(\sigma(t))} \quad\makebox{ and} \quad \ds\sum_{A=1}^k \ds\frac{\partial (\gamma^A_i\circ\sigma)}{\partial t^A}\Big\vert_{t} = -\ds\frac{\partial H}{\partial q^i}\Big\vert_{\gamma(\sigma(t))}\;.
\end{equation}

 Now, we will compute the differential of the function $H\circ\gamma\colon Q\to \R$:
 \begin{equation}\label{d(hgamma)}
d(H \circ \gamma) = (\ds\frac{\partial H}{\partial q^i}\circ \gamma +
(\ds\frac{\partial H}{\partial p^A_j}\circ \gamma) \ds\frac{\partial
\gamma^A_j}{\partial q^i}) \, dq^i\,.
\end{equation}

Then from (\ref{closeness}), (\ref{hegamma}) and (\ref{d(hgamma)}) we obtain
$$\begin{array}{rcl}
d(H\circ\gamma)(\sigma(t))  &=& \left(\ds\frac{\partial H}{\partial q^i}\Big\vert_{\gamma(\sigma(t))} +
\ds\frac{\partial H}{\partial p^A_j}\Big\vert_{\gamma(\sigma(t))} \ds\frac{\partial
\gamma^A_j}{\partial q^i}\Big\vert_{\sigma(t)}\right)dq^i(\sigma(t))
\\\noalign{\medskip}
 &=& \left(-\ds\sum_{A=1}^k \ds\frac{\partial (\gamma^A_i\circ\sigma)}{\partial t^A}\Big\vert_{t} +\ds\frac{\partial \sigma^j}{\partial t^A}\Big\vert_{t}\ds\frac{\partial
\gamma^A_j}{\partial q^i}\Big\vert_{\sigma(t)}\right)dq^i(\sigma(t))
\\\noalign{\medskip}
 &=& \left(-\ds\sum_{A=1}^k \ds\frac{\partial (\gamma^A_i\circ\sigma)}{\partial t^A}\Big\vert_{t} +\ds\frac{\partial \sigma^j}{\partial t^A}\Big\vert_{t}\ds\frac{\partial
\gamma^A_i}{\partial q^j}\Big\vert_{\sigma(t)}\right)dq^i(\sigma(t))
=0\quad .
\end{array}$$
the last term being zero by the chain rule.
 Since $Z$ is integrable, the $k$-vector field $Z^\gamma$ is integrable, then for each point $q\in Q$ we have an integral section $\sigma\colon U_0\subset \R^k\to Q$ of $Z^\gamma$ passing trough    this point,  then
 $$ d(H\circ\gamma)=0\,.$$

 \noindent $(ii)\Rightarrow (i)$

 Let us suppose that $d(H\circ\gamma)=0$ and  $\sigma$ is an integral section of $Z^\gamma$. Now we will prove that $\gamma\circ\sigma$ is a solution to the Hamilton field equations, that is (\ref{hegamma}) is satisfied.

 Since $d(H\circ\gamma)=0$, from (\ref{d(hgamma)}) we obtain
 \begin{equation}\label{d(hgamma)=0}
 0= \ds\frac{\partial H}{\partial q^i}\circ \gamma +
(\ds\frac{\partial H}{\partial p^A_j}\circ \gamma) \ds\frac{\partial
\gamma^A_j}{\partial q^i}\;.
\end{equation}

From (\ref{localhameq}) and (\ref{zgamma}) we know that
$$
  Z^\gamma_A = (\ds\frac{\partial H}{\partial p^A_i}\circ\gamma) \ds\frac{\partial}{\partial q^i}
$$
and then since $\sigma$ is an integral section of $Z^\gamma$ we obtain
\begin{equation}\label{sigmasi}
\ds\frac{\partial \sigma^i}{\partial t^A} = \ds\frac{\partial H}{\partial p^A_i}\circ\gamma\circ\sigma\;.
\end{equation}

On the other hand, from (\ref{closeness}), (\ref{d(hgamma)=0}) and (\ref{sigmasi}) we obtain
$$\begin{array}{lcl}
\ds\sum_{A=1}^k\ds\frac{\partial (\gamma^A_i\circ \sigma)}{\partial t^A} &=& \ds\sum_{A=1}^k (\ds\frac{\partial  \gamma^A_i}{\partial q^j}\circ\sigma) \ds\frac{\partial \sigma^j}{\partial t^A} = \ds\sum_{A=1}^k (\ds\frac{\partial  \gamma^A_i}{\partial q^j}\circ\sigma)(\ds\frac{\partial H}{\partial p^A_j}\circ\gamma\circ\sigma)
\\\noalign{\medskip} &=&
\ds\sum_{A=1}^k(\ds\frac{\partial  \gamma^A_j}{\partial q^i}\circ\sigma)(\ds\frac{\partial H}{\partial p^A_j}\circ\gamma\circ\sigma) = - \ds\frac{\partial H}{\partial q^i}\circ\gamma\circ\sigma\;.
\end{array}$$
and thus we have proved that $\gamma\circ\sigma $ is a solution to the Hamilton field equations.

\qed

\begin{remark}{\rm In the particular case $k=1$ the above theorem can be found in \cite{lmm2}.}\end{remark}

\begin{theorem}\label{hamjacobi1}
Let $Z$ be a solution of the Hamilton equations (\ref{hameq}) and
$\gamma : Q \longrightarrow (T^1_k)^*Q$ be a   closed section  of
$\Pi_Q : (T^1_k)^*Q \longrightarrow Q$, that is, $\gamma =
(\gamma^1, \dots, \gamma^k)$ where   each $\gamma^A$ is an ordinary
closed 1-form on $Q$. Then, the following statements are equivalent:
\begin{enumerate}
\item  $Z\circ \gamma - T^1_k\gamma(Z^\gamma) \in \ker \flat$
\item $ d(H \circ \gamma) = 0$.
\end{enumerate}
\end{theorem}
\proof We know that if $Z_A$ and $\gamma^A$ are locally given by
$$ Z_A= Z_A^i\ds\frac{\partial }{\partial q^i} + (Z_A)^B_i\ds\frac{\partial}{\partial p^B_i} \quad , \quad \gamma^A= \gamma^A_idq^i\;.$$ then
$Z^\gamma_A= (Z_A^i\circ\gamma)\ds\frac{\partial}{\partial q^i}$. Thus   a direct computation  shows that
$Z\circ \gamma - T^1_k\gamma(Z^\gamma) \in \ker \flat$ is locally written as
\begin{equation}\label{locexpr}
\left( (Z_A)^B_i\circ\gamma - (Z_ A^j\circ\gamma ) \ds\frac{\partial \gamma^B_ i}{\partial q^j}
\right)\left(\ds\frac{\partial}{\partial p^B_i}\circ\gamma\right)=(Y_A)^B_i\circ \gamma\left(\ds\frac{\partial}{\partial p^B_i}\circ\gamma\right)\;.
\end{equation}
where $(Y_A)^A_i=0$.

Now, we are  ready to prove the result.

Assume that $(i)$ holds, then from (\ref{localhameq}), (\ref{kerb}) and (\ref{locexpr}) we obtain that
\[\begin{array}{lcl}
0 &=& \ds\sum_{A=1}^k\left( (Z_A)^A_i\circ\gamma - (Z_ A^j\circ\gamma ) \ds\frac{\partial \gamma^A_ i}{\partial q^j}\right)
 \\\noalign{\medskip}  &=& -\left(
 (\ds\frac{\partial H}{\partial q^i}\circ\gamma) + (\ds\frac{\partial H}{\partial p^A_ j}\circ\gamma)\ds\frac{\partial \gamma^A_i}{\partial q^j}\right)
\\\noalign{\medskip} &=& - \left(
(\ds\frac{\partial H}{\partial q^i}\circ\gamma) + (\ds\frac{\partial
H}{\partial p^A_ j}\circ\gamma)\ds\frac{\partial
\gamma^A_j}{\partial q^i}\right)
\end{array}\]
where in the last identity we are using the closeness of $\gamma$ (see (\ref{closeness})).
Therefore, $d(H\circ\gamma)=0$ (see (\ref{d(hgamma)})).

The converse is proved in a similar way by reversing the arguments.

\qed
\begin{remark}{\rm In the particular case $k=1$ the above theorem can be found in \cite{lmm2}.}\end{remark}

\begin{remark}
{\rm It should be noticed that if $Z$ and $Z^\gamma$ are
$\gamma$-related, that is, $Z_A= T\gamma(Z^\gamma_A)$, then $d(H\circ \gamma) = 0$, but the converse does
not hold. }
\end{remark}

\begin{corollary}
Let $Z$ be a solution of (\ref{hameq}), and $\gamma$ a closed
section of $\Pi_Q : (T^1_k)^*Q \longrightarrow Q$, as in the above
theorem.  If $Z$ is integrable then the following statements are
equivalent:
\begin{enumerate}
\item  $Z\circ \gamma - T^1_k\gamma(Z^\gamma) \in \ker \flat$;
\item $ d(H \circ \gamma) = 0$;
\item If $\sigma\colon U\subset \R^k\to Q$ is an integral section of $Z^\gamma$  then $\gamma\circ\sigma$ is a solution of the Hamilton equations.
\end{enumerate}
 \end{corollary}

The equation
\begin{equation}\label{hjeq}
d(H\circ\gamma)=0
\end{equation}
can be considered as the geometric version of the Hamilton-Jacobi equation for $k$-symplectic field theories. Notice that in local coordinates, equation (\ref{hjeq}) reads us
\[
H(q^i,\gamma^A_i(q))=constant\;.
\]
which when $\gamma^A=dW^A$, where $W^A\colon Q\to \R$ is a function, takes the more familiar form
\[
H(q^i,\ds\frac{\partial W^A}{\partial q^i})=constant\,.
\]

\section{ The Hamilton-Jacobi problem and the $k$-symplectic Lagrangian
 field theory}

\subsection{$k$-symplectic Lagrangian  field theory}

Consider now the Lagrangian formalism. Let $L\in
\mathcal{C}^{\infty}(T^1_kQ)$ be a regular Lagrangian function, that is, the Hessian matrix $\ds(\ds\frac{\partial^2 L}{\partial v^i_A\partial v^j_B})$ has maximal rank. We
can endow $T^1_kQ$ with a $k$-symplectic structure given by the
family of $k$ $2$-forms $(\omega_L^1,\ldots, \omega_L^k)$, where
$$
\omega_L^A=FL^*(\omega^A)
$$
and $FL:T^1_kQ\to (T^1_k)^*Q$ is the Legendre transformation
introduced by G\"{u}nther \cite{gun}. In local coordinates
$FL(q^i,v^i_A)=(q^i,\ds\frac{\partial L}{\partial v^i_A})$. Thus we
have
$$
\omega_L^A=dq^i\wedge d(\ds\frac{\partial L}{\partial v^i_A})\;.
$$
Denote also by $\theta_L^A=FL^*(\theta^A)$.

We define the Lagrangian energy function as $E_L =\Delta(L)-L$ where $\Delta\in\mathfrak{X}(T^1_kQ)$ is the Liouville vector field, that is, the infinitesimal generator of the flow
$$
\psi\colon\R\times T^1_kQ\to T^1_k Q\quad ,\quad \psi(s, {v_1}_q,\ldots, {v_k}_q)=(e^s{v_1}_q,\ldots, e^s{v_k}_q)\,.
$$

As in the Hamiltonian formalism, we consider the mapping
\begin{eqnarray*}
\flat_L &:& T_k^1(T_k^1Q) \longrightarrow T^*(T_k^1Q)\\
&& z = (z_1, \dots, z_k)  \mapsto \flat_L(z) = \hbox{trace} \left(
i_{\ds z_A} \, \omega_L^B\right)=\sum_{A=1}^ki_{\ds z_A}\omega_L^A
\underline{}\end{eqnarray*}

First we study the kernel of $\flat_L$.

Let $Z=(Z_l,\ldots, Z_{k})$ be a
$k$-vector field on $T^1_kQ$, that is, each $Z_{A}$ is a
vector field on $T^1_kQ$ locally given
$$
Z_A=Z_A^i\ds\frac{\partial }{\partial q^i} +
(Z_A)^i_B\ds\frac{\partial}{\partial v^i_B}
$$
then
\begin{equation}\label{loceleq}
\begin{array}{lcl}
\flat_L(Z_1,\ldots, Z_{k}) &=&
\left[(\displaystyle\ds\frac{\partial^2 L}{\partial q^i\partial
v^j_A}\,-\,\ds\frac{\partial^2 L}{\partial q^j\partial v^i_A})Z^j_A -
\ds\frac{\partial^2 L}{\partial v^j_B\partial v^i_A}(Z_A)^j_B \right] dq^i
\\\noalign{\medskip} & + & \displaystyle
\ds\frac{\partial^2 L}{\partial v^j_B\partial v^i_A}\,Z_A^i\,dv^j_B\;.
\end{array}
\end{equation}

Therefore, since $L$ is regular, $(Z_1,\ldots, Z_{k})\in \ker
\flat_L$ if and only if
\begin{equation}\label{kerbl}
Z^i_A=0 \quad , \quad  \ds\frac{\partial^2 L}{\partial v^j_B\partial v^i_A}(Z_A)^j_B =0\;.
\end{equation}

Now, we look for the solutions of the equation
\begin{equation}\label{eleq}
\flat_L(Z)=dE_L
\end{equation}
which is locally expressed as follows:
\begin{equation}\label{locELeq}
Z_A^i=v^i_A\quad ,\quad\ds\frac{\partial^2 L}{\partial q^j\partial
v^i_A}v^j_A+ \ds\frac{\partial^2 L}{\partial
v^j_B\partial v^i_A}\,(Z_A)^j_B =\ds\frac{\partial L}{\partial q^i}\;.
\end{equation}

Then, if $Z$ is integrable, an integral section
$$
\sigma(t^1, \dots, t^k) = (q^i(t^1, \dots, t^k), v^i_A(t^1, \dots,
t^k))
$$
satisfies the Euler-Lagrange equations
\begin{equation}\label{ELeq}
v^i_A=\ds\frac{\partial q^i}{\partial t^A}\quad , \quad
\displaystyle\sum_{A=1}^k\ds\frac{d}{dt^A}\left(\ds\frac{\partial
L}{\partial v^i_A}\right)  = \ds\frac{\partial L}{\partial q^i}\;.
\end{equation}

\subsection{Hamilton-Jacobi problem on $T^1_kQ$}\

In this section we formulate the Hamilton-Jacobi problem on the tangent bundle of $k^1$-velocities.

In the section \ref{HJproblem} we comment that the Hamilton-Jacobi problem in the $k$-symplectic framework consists in finding $k$ functions $W^1,\ldots, W^k\colon Q\to \R$ such that
\begin{equation}\label{HJ-ksym2}
H(q^i,\ds\frac{\partial W^1}{\partial q^i}, \ldots
,\ds\frac{\partial W^k}{\partial q^i}) = constant
\end{equation}

In a geometric terms, equation (\ref{HJ-ksym2}) can be written as $H\circ (dW^1,\ldots, dW^k)=constant$, where $(dW^1, \ldots, dW^k)$ is
a section of the tangent bundle of $k^1$-covelocities, $(T^1_k)^*Q$. As we have seen in the section \ref{HJproblem},
we look for a closed section $\gamma=(\gamma^1,\ldots,\gamma^k)$ of $\Pi_Q$ such that $H\circ\gamma=\gamma^*H=constant$. Let us observe that the section $\gamma$ is closed, and hence locally exact, $\gamma^A=dW^A$.
 The condition $d\gamma=0$ can be alternatively be expressed in terms of the canonical forms $(\omega^1,\ldots,\omega^k)$ in the form $\gamma^*\omega^A=0,\; A=1\ldots, k$, so that one
 can reformulate the Hamilton-Jacobi geometric problem in the form: {\it find a section $\gamma=(\gamma^1,\ldots, \gamma^k)\colon Q\to (T^1_k)^*Q$ of $\Pi_Q$ such that
\begin{equation}\label{altHJproblem}
\gamma^*H=constant \quad , \quad \gamma^*\omega^A=0\,,\quad A=1,\ldots, k\,.
\end{equation}}

Consider now the Lagrangian $k$-symplectic formalism. Let $L\in \mathcal{C}^\infty(T^1_kQ)$ be a regular Lagrangian function and $\theta_L^A,\omega_L^A,\, A=1,\ldots, k$ the associated Lagrangian forms. A literal translation of the above formulation of the Hamilton-Jacobi problem for the tangent bundle of $k^1$-covelocities to the tangent bundle of $k^1$-velocities would be: {\it to find a section $X=(X_1,\ldots, X_k)\colon Q\to T^1_kQ$ of $\tau$ such that
\begin{equation}\label{altHJproblemLang}
X^*E_L=constant \quad , \quad X^*\omega_L^A=0\,,\quad A=1,\ldots, k\,,
\end{equation}
where $E_L$ denotes the energy function associated to $L$}. The last family of conditions $X^*\omega_L^A=0,\; A=1,\ldots, k$ implies that the section $X$ is associated (at least locally) with a mapping $W=(W^1,\ldots, W^k)\colon Q\to \R^k$ by means of the relation $X^*\theta_L^A= dW^A$. In fact,
$$
0=X^*\omega_L^A=-d(X^*\theta_L^A)\,
$$
then, the $1$-form $X^*\theta_L^A$ is closed and therefore locally exact, thus there is a function $W^A\colon Q\to \R$ defined on
a neighborhood of each point of $Q$, such that, $X^*\theta_L^A=dW^A$. Locally, this means
$$
\ds\frac{\partial L}{\partial v^i_A}\circ X =\ds\frac{\partial W^A}{\partial q^i}\,.
$$

In this section we will give a geometric version of the Hamilton-Jacobi equation (\ref{altHJproblemLang}).

\begin{theorem}
Let $X=(X_1,\ldots, X_k)$  be an integrable $k$-vector field on $Q$ such that $X^*\omega_L^A=0$.   Then, the following statements are equivalent:
\begin{enumerate}
\item If $\sigma\colon U\subset \R^k\to Q$ is an integral section of $X$  then $\sigma^{(1)}$ is a solution of the Euler-Lagrange equations;

\item $d(E_L\circ X)=0$.

\end{enumerate}
\end{theorem}

\proof Since $X^*\omega_L^A=0$ we have that the $1$-forms $X^*\theta_L^A$ are closed. The closeness of this $1$-forms states that
\begin{equation}\label{closenesslag}
\ds\frac{\partial^2 L}{\partial q^j\partial v^i_A}\circ X + \left(\ds\frac{\partial^2 L}{\partial v^k_B\partial v^i_A}\circ X\right)\ds\frac{\partial X^k_B}{\partial q^j} = \ds\frac{\partial^2 L}{\partial q^i \partial v^j_A}\circ X + \left(\ds\frac{\partial^2 L}{\partial v^k_B\partial v^j_A}\circ X\right)\ds\frac{\partial X^k_B}{\partial q^i}
\end{equation}
where $X(q)=(q,X^k_B(q))$.

In first place, let us suppose that $\sigma(t)=(\sigma^i(t))$ is an integral section of $X$
\begin{equation}\label{intsect}
X^i_A\circ\sigma=\ds\frac{\partial \sigma^i}{\partial t^A}\,,
\end{equation}
such that $\sigma^{(1)}$ is a solution of the Euler-Lagrange equations.

We will prove that $d(E_L\circ X)=0$ along $\sigma$.
Now, we   compute the differential of the function $E_L\circ X\colon Q\to \R$. In local coordinates we obtain that
$$
E_L\circ X=X^i_A(\ds\frac{\partial L}{\partial v^i_A}\circ X) - L\circ X
$$
then
\begin{equation}\label{d(ElX)}
d(E_L\circ X)= \left(X^i_A\left(\ds\frac{\partial^2 L}{\partial q^j\partial v^i_A}\circ X + (\ds\frac{\partial^2 L}{\partial v^k_B\partial v^i_A}\circ X)\ds\frac{\partial X^k_B}{\partial q^j}\right) - \ds\frac{\partial L}{\partial q^j}\circ X\right)dq^j\; .
\end{equation}

Therefore, from (\ref{ELeq}), (\ref{closenesslag}), (\ref{intsect}) and (\ref{d(ElX)}) we obtain

{\small\[\begin{array}{ll}
&d(E_L\circ X)(\sigma(t))
\\\noalign{\medskip}
=& \left(X^i_A(\sigma(t)) (\ds\frac{\partial^2 L}{\partial q^j\partial v^i_A}\Big\vert_{X(\sigma(t))} + \ds\frac{\partial^2 L}{\partial v^k_B\partial v^i_A}\Big\vert_{X(\sigma(t))}\ds\frac{\partial X^k_B}{\partial q^j}\Big\vert_{\sigma(t)} ) - \ds\frac{\partial L}{\partial q^j}\Big\vert_{X(\sigma(t))}\right)dq^j(\sigma(t))
\\\noalign{\medskip}
=& \left(X^i_A(\sigma(t)) (\ds\frac{\partial^2 L}{\partial q^i\partial v^j_A}\Big\vert_{X(\sigma(t))} + \ds\frac{\partial^2 L}{\partial v^k_B\partial v^j_A}\Big\vert_{X(\sigma(t))}\ds\frac{\partial X^k_B}{\partial q^i}\Big\vert_{ \sigma(t)}\right) - \ds\frac{\partial L}{\partial q^j}\Big\vert_{X(\sigma(t))} )dq^j(\sigma(t))
\\\noalign{\medskip}
=& \left(\ds\frac{\partial \sigma^i}{\partial t^A}\Big\vert_{t} \ds\frac{\partial^2 L}{\partial q^i\partial v^j_A}\Big\vert_{ \sigma^{(1)}(t)} + \ds\frac{\partial^2\sigma^k}{\partial t^A\partial t^B}\Big\vert_{t}\ds\frac{\partial^2 L}{\partial v^k_B\partial v^j_A}\Big\vert_{ \sigma^{(1)}(t)}  - \ds\frac{\partial L}{\partial q^j}\Big\vert_{ \sigma^{(1)}(t)}\right)dq^j(\sigma(t))=0
\end{array}\]}

Thus $d(E_L\circ X)=0$ along $ \sigma$ and since the $k$-vector field $X$ is integrable, for each point $q\in Q$ we have an integral section $\sigma$ of $X$ passing trough this point, then
$$d(E_L\circ X)=0\,.$$

The converse is proved in a similar way by reversing the arguments.
\qed

\begin{theorem}\label{hamjacobi11}
Let $Z$ be a solution of the Euler-Lagrange equations (\ref{eleq}) and
$X : Q \longrightarrow  T^1_k Q$ be a     section  of
$\tau : T^1_kQ \longrightarrow Q$, such that $X^*\omega_L^A=0$. Then, the following statements are equivalent:
\begin{enumerate}
\item  $Z\circ X - T^1_kX(X) \in \ker \flat_L$
\item $ d(E_L \circ X) = 0$.
\end{enumerate}
\end{theorem}
\proof A direct computation shows that if $Z_A$ and $X_A$ are locally given by
$$
 Z_A= v^i_A\ds\frac{\partial }{\partial q^i} + (Z_A)^i_B\ds\frac{\partial}{\partial v^i_B} \quad , \quad X_A= X_A^i\ds\frac{\partial}{\partial q^i}
$$
then
\begin{equation}\label{zx}
Z_A\circ X - TX(X_A)= \left((Z_A)^j_B\circ X - X^i_A\ds\frac{\partial X^j_B}{\partial q^i} \right)(\ds\frac{\partial}{\partial v^j_B}\circ X)
\end{equation}

Now, we are  prepared to prove the result.

Assume that $(i)$ holds, then from (\ref{kerbl}), (\ref{locELeq}) and (\ref{zx}) we obtain that
\[\begin{array}{lcl}
0 &=&  \left((Z_A)^j_B\circ X - X^k_A\ds\frac{\partial X^j_B}{\partial q^k} \right)(\ds\frac{\partial^2L}{\partial v^j_B\partial v^i_A}\circ X)
\\\noalign{\medskip} &=&
\ds\frac{\partial L}{\partial q^i}\circ X -
X^j_A\left(\ds\frac{\partial^2 L}{\partial q^j\partial v^i_A}\circ
X+(\ds\frac{\partial^2 L}{\partial v^k_B\partial v^i_A}\circ
X)\ds\frac{\partial X^k_B}{\partial q^j}\right)
\\\noalign{\medskip} &=&
\ds\frac{\partial L}{\partial q^i}\circ X -
X^j_A\left(\ds\frac{\partial^2 L}{\partial q^i\partial v^j_A}\circ
X+(\ds\frac{\partial^2 L}{\partial v^k_B\partial v^j_A}\circ
X)\ds\frac{\partial X^k_B}{\partial q^i}\right)
\end{array}\]
where in the last identity we are using the closeness of $X^*\theta_L^A$ (see (\ref{closenesslag})).
Therefore, $d(E_L\circ X)=0$ (see (\ref{d(ElX)})).

The converse is proved in a similar way by reversing the arguments.

\qed

\begin{corollary}
Let $Z$ be a solution of (\ref{eleq}), and
$X$ an integrable $k$-vector field on $Q$ such that $X^*\omega_L^A=0$. Then the following statements are equivalent:
\begin{enumerate}
\item  $Z\circ X - T^1_kX(X) \in \ker \flat_L$;
\item $ d(E_L \circ X) = 0$;
\item If $\sigma\colon U\subset \R^k\to Q$ is an integral section of $X$  then $\sigma^{(1)}$ is a solution of the Euler-Lagrange equations.
\end{enumerate}
 \end{corollary}

The equation
\begin{equation}\label{LH-Jeq}
d(E_L\circ { X})=0\,
\end{equation}
can be considered as  the geometric Lagrangian version of the Hamilton-Jacobi equation for $k$-symplectic field theories.  Notice that in
local coordinates, the equation (\ref{LH-Jeq}) reads us
\begin{equation}\label{elx}
E_L(q^i,X^i_A(q^j))=constant\,.
\end{equation}

If $X^*\omega_L^A=0$, then
$0=X^*\omega_L^A=-X^*d\theta_L^A=-d(X^*\theta_L^A)$, we have that
every point has an open neighborhood $U\subset Q$ where there exists
$k$ functions $W^A\in\mathcal{C}^\infty(U)$ such that
$$dW^A=X^*\theta_L^A= (\ds\frac{\partial L}{\partial
v_A^i}\circ X) dq^i$$ and then in local coordinates this means
$$
 \ds\frac{\partial L}{\partial
v_A^i}\circ X =  \ds\frac{\partial W^A}{\partial
q^i}\,.
$$

If the Lagrangian $L$ is regular, then the Legendre
transformation $FL$ is a local diffeomorphism, then in a
neighborhood of each point of $T^1_kQ$ we have $H=E_L\circ FL^{-1}$.
Therefore, if we consider the section $\gamma=(X^*\theta_L^1,\ldots, X^*\theta_L^k): Q \to (T^1_k)^*Q$, locally given by
$$
\gamma(q)=(q ,\ds\frac{\partial W^A}{\partial q^i}\Big\vert_{q})
$$ we have
$$
\begin{array}{lcl}
 H(\gamma(q)) &= &\displaystyle H(q^i,\ds\frac{\partial W^A}{\partial
q^i})= H(q^i,\ds\frac{\partial L}{\partial v^i_A}\Big\vert_{X(q)}) =
H\circ FL(X(q))\\\noalign{\medskip}
 &=&\displaystyle E_L(X(q))=E_L(q^i,X^i_A(q^B))=constant
\end{array}
$$
and thus (\ref{elx}) takes the form
$$
H(q^i,\ds\frac{\partial W^A}{\partial q^i})=constant\,,
$$
where $H=E_L\circ FL^{-1}$ .

\section{Example}

\textbf{Vibrating string.} In this example we consider the theory of a
vibrating string. Coordinates $(t^1,t^2)$  are interpreted as the
time and the distance along the string, respectively.

Let us denote by $(q,p^1,p^2)$ the coordinates of $(T^1_2)^*\R$ and let us consider
the Hamiltonian

$$
\begin{array}{lccl}
H \colon & (T^1_2)^*\R & \to & \R\\\noalign{\medskip}
 & (q,p^1,p^2) & \mapsto & \ds\frac{1}{2}\left(\ds\frac{(p^1)^2}{\sigma}-\ds\frac{(p^2)^2}{\tau}\right)
\end{array}
$$ where $\sigma$ and $\tau$ are certain constants of the mechanical
system. In a real string, these constants represent the linear mass density, that is,
a measure of mass per unit of length and Young's module of the system related to the
tension of the string, respectively.
%, see for instance \cite{god}.

Let $\gamma\colon \R\to (T^1_2)^*\R $ be the  section of $\pi_\R$ defined by $\gamma(q)=(aq\,dq,bq\,dq)$
where $a$ and $b$ are two constants such that $\tau a^2=\sigma b^2$. This section $ \gamma$ satisfies
the condition $d(H\circ \gamma)=0$, therefore, the condition $(i)$ of the Theorem \ref{hjth} holds.

The $2$-vector field $Z^\gamma=(Z^\gamma_1,Z^\gamma_2)$ is locally given by
$$
Z^\gamma_1=\ds\frac{a}{\sigma} q\ds\frac{\partial}{\partial q} \quad , \quad Z^\gamma_2=-\ds\frac{b}{\tau} q\ds\frac{\partial}{\partial q}
$$

If $\psi\colon \R^2\to \R$ is an integral section of $Z^\gamma$, then
$$\ds\frac{\partial \psi}{\partial t^1}=\ds\frac{a}{\sigma}\psi \quad , \quad
\ds\frac{\partial \psi}{\partial t^2}=-\ds\frac{b}{\tau}\psi,$$ thus
$$\psi(t^1,t^2)= C\,\hbox{exp }\left({\ds\frac{a}{\sigma}t^1-\ds\frac{b}{\tau}t^2 }\right),\quad C\in \R$$

By Theorem \ref{hjth} one obtains that the map $
\phi=\gamma\circ \psi$, locally given by $$(t^1,t^2) \mapsto (\psi(t^1,t^2),a\psi(t^1,t^2),b\psi(t^1,t^2)),$$
is a solution of the Hamilton equations associated to $H$, that is,
$$\begin{array}{ccl}
0 &=& a\ds\frac{\partial \psi}{\partial t^1} + b\ds\frac{\partial \psi}{\partial t^2} \\\noalign{\medskip}
\ds\frac{a}{\sigma}\psi &=& \ds\frac{\partial \psi}{\partial t^1}
\\\noalign{\medskip}
-\ds\frac{b}{\tau}\psi &=& \ds\frac{\partial \psi}{\partial t^2}\end{array}
$$

Let us observe that from this system one obtains that $\psi$ is a solution of the motion equation of the vibrating string, that is,
\begin{equation}\label{ec42110}
\sigma\partial_{11} \psi -\tau \partial_{22}\psi=0,
\end{equation} where $\psi(t^1,t^2)$ denotes the displacement of each point of the
string as function of the time $t^1$ and the position $t^2$.

\section*{Acknowledgments}
We acknowledge the partial financial support of {\sl Ministerio de
Innovaci\'{o}n y Ciencia}, Project MTM2007-62478, MTM2008-00689,
MTM2008-03606-E/, MTM2009-13383 and project Ingenio Mathematica(i-MATH) No.
CSD2006-00032 (Consolider-Ingenio2010).

\end{document}